\documentclass[conference,a4paper]{IEEEtran}
\IEEEoverridecommandlockouts

\usepackage[hidelinks]{hyperref}
\usepackage[cmex10]{amsmath}
\usepackage{amssymb,amsfonts}
\usepackage{dblfloatfix}

\usepackage[ruled,vlined]{algorithm2e}
\usepackage{graphicx}
\graphicspath{{Figures/PDF/}{Figures/PNG/}}

\usepackage{booktabs}
\usepackage{siunitx}
\usepackage[numbers,compress]{natbib}
\usepackage{texnames}
\usepackage{placeins}
\usepackage{bm,bbm}
\usepackage{orcidlink}

\begin{document}

\title{LEARNING-BASED 3D RECONSTRUCTION OF POWER NETWORKS FROM AERIAL POINT CLOUDS%
\thanks{Portions of this work are covered by a pending patent application.}}

\author{
\IEEEauthorblockN{Rishabh Jain}
\IEEEauthorblockA{\textit{AiDash}\\
rishabh.jain@aidash.com}
\and
\IEEEauthorblockN{Anuja Saini}
\IEEEauthorblockA{\textit{AiDash}\\
anuja.saini@aidash.com}
\and
\IEEEauthorblockN{Vishal Jain}
\IEEEauthorblockA{\textit{AiDash}\\
vishal@aidash.com}
}

\maketitle

\begin{abstract}
This paper presents an end-to-end framework for reconstructing overhead power utility network topology and extracting span-level physical metadata from large-scale aerial LiDAR. The pipeline begins with semantic segmentation of the input point cloud using an improved KPConv-based model \cite{thomas2019kpconv}, in which data sampling and loss functions are adapted to emphasize pole and conductor (wire) classes. Network topology inference then proceeds in two stages: (i) pole instances are obtained by clustering pole-class points and validating candidates using geometric criteria, including height and verticality estimated via PCA \cite{jolliffe2002principal}, and (ii) candidate pole pairs are evaluated using a heuristic method and a lightweight ResNet-based classifier \cite{he2016resnet} on 2D top view projections of pole and wire point distributions to determine whether a physical conductor span exists. By explicitly classifying candidate spans, the approach mitigates common failure modes of heuristic connectivity rules in dense or cluttered scenes and under partial wire observation. For each validated wire, attributes regarding utility infrastructure geometry are computed, including endpoint conductor heights, ground elevation, sag-related lowest-point features, conductor arrangement, and wire width. Evaluation on multiple real-world aerial LiDAR datasets demonstrates decimeter-level endpoint height accuracy and approximately 9\% relative improvement of recall in topology reconstruction performance compared to heuristic nearest-neighbor baselines, with larger gains in complex layouts.
\end{abstract}

\begin{IEEEkeywords}
Aerial LiDAR, utility infrastructure mapping, utility pole detection, semantic segmentation, powerline detection, KPConv, ResNet, Deep learning, geospatial analytics, GIS
\end{IEEEkeywords}

\section{Introduction}


Digital representations of overhead distribution networks are central to asset inventory, vegetation clearance assessment, and asset management. Aerial LiDAR provides direct 3D measurements of conductors and supporting structures at scale; however, converting point clouds into a \emph{topologically correct} network graph remains difficult due to occlusions, vegetation clutter, heterogeneous point density, and intermittent wire returns \cite{powerline_lidar_review,individual_wire_extraction}.

Operational pipelines often rely on manual effort or heuristic connectivity rules, which can mis-connect poles in dense layouts or fail under sparse wire observations \cite{isprs2019_utility_poles,powerline_lidar_review}.
 The framework described in this paper addresses these limitations using a learning-assisted approach that combines robust pole extraction with span classification based on compact 2D pole–wire evidence maps.


The main contributions of this work are:
\begin{itemize}
	\item An end-to-end pipeline for extracting poles and conductor spans from aerial LiDAR, including a KPConv-based semantic segmentation stage \cite{thomas2019kpconv} when labels are unavailable.
    \item A lightweight ResNet-based span classifier \cite{he2016resnet} validates candidate pole pairs using compact 2D pole–wire projections, improving topology inference under clutter and sparse wire observations.

	\item A span-level metadata extraction module that estimates conductor endpoint heights, ground elevation, sag-related features, conductor arrangement, and wire width.
\end{itemize}


\section{Related Work}

Extraction of utility poles and overhead power lines from LiDAR point clouds has been studied across airborne (ALS), mobile (MLS/MTLS), and terrestrial (TLS) platforms. Many approaches combine corridor selection, geometric filtering, and rule-based processing to isolate pole-like and wire-like structures from surrounding clutter \cite{isprs2019_utility_poles,csit_powerline_detection}. Despite this progress, robustness remains limited under vegetation occlusion, uneven point density, sparse/intermittent wire returns, and varying acquisition geometry \cite{powerline_lidar_review}.

Pole detection is typically formulated as extracting pole-shaped objects \cite{road_lidar_poles,fig2017_lidar_classification}. While effective for isolated poles, these methods can produce false positives in dense urban/roadside scenes where trees, signs, or building edges exhibit similar vertical structure \cite{csit_powerline_detection}.



Network topology reconstruction from LiDAR relies on continuity cues and elevation consistency within predefined pole-to-pole corridors. Existing approaches model individual wires or wire bundles using curve fitting, local smoothness constraints, or parametric representations of linear or catenary structures, and infer connectivity by linking nearby poles based on distance and angular heuristics, validated through corridor-based wire evidence \cite{individual_wire_extraction,isprs2019_utility_poles,fig2017_lidar_classification}. While computationally efficient, this heuristic formulation degrades under sparse, fragmented, or missing wire observations and is particularly prone to mis-connections in dense layouts and at junctions \cite{powerline_lidar_review}. \autoref{fig:missing_wire_examples} shows a sample of intermittent wire observations.


\section{Problem Statement and Assumptions}
\subsection{Problem Statement}
Given large-scale aerial LiDAR point clouds, the goal is to reconstruct a topologically correct network graph by inferring physical connectivity of overhead conductors between utility poles, and to estimate span-level geometric attributes for engineering analysis and GIS use.

\subsection{Assumptions}
The framework assumes: (i) sufficient point density to observe poles and conductors with reasonable continuity (mentioned in \autoref{tab:datasets}); (ii) each reconstructed edge corresponds to a single physical conductor span supported by exactly two poles (i.e., no intermediate supports within a span).

\subsection{Data Specifications}
\textbf{Input:} aerial LiDAR point clouds in LAS/LAZ format.\\
\textbf{Outputs:} (a) semantically labeled point cloud (pole, wire, ground, other); (b) validated pole instances with estimated locations and heights; (c) wire-span topology exports (e.g., shapefiles per tile/area); and (d) span-level physical metadata as attributes and/or per-span GeoJSON.

\section{Methodology}
\label{sec:method}
\begin{figure}[ht]
	\centering
	\includegraphics[width=0.8 \linewidth]{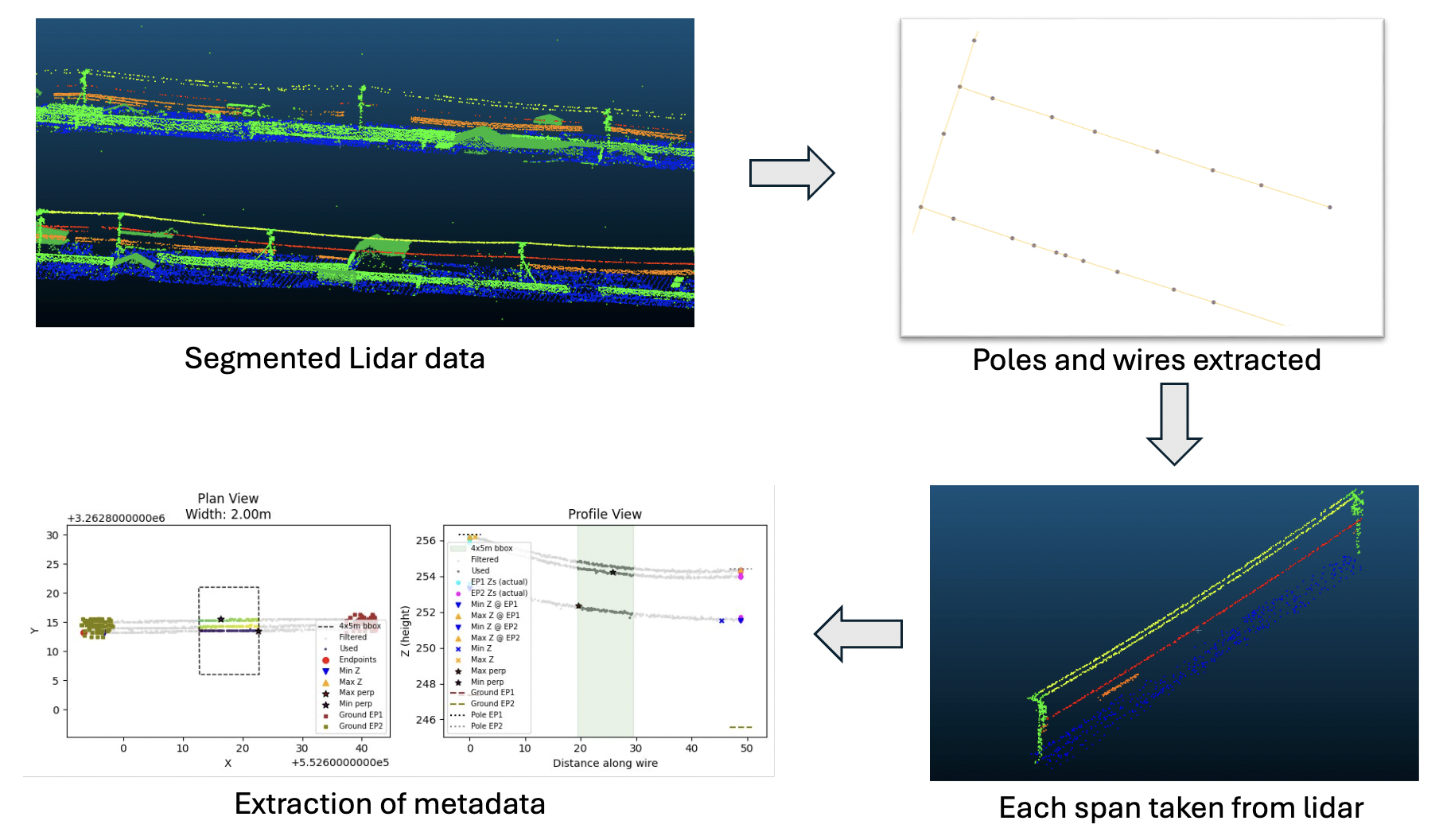}
	\caption{Workflow depiction of separate stages}\label{fig:outerbound}
\end{figure}

The end-to-end workflow (\autoref{fig:outerbound}) consists of:
\begin{enumerate}
	\item \textbf{Semantic segmentation:} if labels are missing, a deep learning model assigns pole, wire, ground, and surrounding classes.
	\item \textbf{Network topology extraction:} poles are detected; candidate spans are generated and validated via a heuristic method and a model-based span detection.
	\item \textbf{Wire metadata extraction:} LiDAR is cropped per span and used to compute endpoint heights, sag features, arrangement, and wire width.
\end{enumerate}


\subsection{Stage 1: Semantic Segmentation} 
\label{sec:seg}
We adopt a KPConv-based semantic segmentation backbone \cite{thomas2019kpconv} and extend it into an improved KPConvNet tailored for thin, vertically oriented utility structures in aerial LiDAR. Each point is augmented with local geometric descriptors (linearity, planarity, scattering, verticality, normalized height, local density, and a z-variance ratio) computed efficiently from neighborhood statistics without per-point eigen-decomposition, and contextual reasoning is strengthened using local multi-head attention over kNN neighborhoods (O(N·k)) together with residual KPConv blocks initialized via Fibonacci-sphere kernels. Multiscale feature fusion concatenates encoder features across resolutions to preserve fine wire/pole boundary cues. Training uses an equal-weight composite of Focal Loss ($\gamma=2.0$) and Lovász-Softmax with class weights $\sqrt{1/f_i}$ capped at 25, and is accelerated with Distributed Data Parallel training and mixed precision (AMP); the key novelty is combining the efficient geometric feature augmentation with localized attention and multiscale KPConv fusion under an imbalance-aware, IoU-focused objective to improve pole/wire segmentation.

\subsection{Stage 2: Network Topology Extraction}


\subsubsection{Pole Detection}
Pole-class points are clustered in the XY plane using DBSCAN \cite{ester1996density} and validated using geometric criteria (\autoref{alg:pole_detection}): height, PCA-based verticality \cite{jolliffe2002principal}, and aspect ratio.

\begin{algorithm}[ht]
\caption{Pole Detection from Point Clouds}
\label{alg:pole_detection}
\KwIn{Point cloud $\mathcal{P}$ with semantic labels}
\KwOut{Validated poles $\mathcal{V}$}

$\mathcal{P}_{p}\leftarrow\{p\in\mathcal{P}\mid y(p)=\text{pole}\}$;\;
$\mathcal{V}\leftarrow\emptyset$;\;

\ForEach{$C\in \mathrm{DBSCAN}_{xy}(\mathcal{P}_{p};\ \epsilon = 1\,\mathrm{m},\ \text{minPts}=10)$}{
$h\leftarrow z_{\max}(C)-z_{\min}(C);\; w\leftarrow \max(\sigma_x(C),\sigma_y(C));\; \alpha\leftarrow\left|\mathbf{v}_1(C)\cdot\hat{\mathbf{z}}\right|$;\;
\lIf{$h\ge4\,\mathrm{m}\ \land\ \alpha\ge0.8\ \land\ h>3w$}
{$\mathcal{V}\leftarrow \mathcal{V}\cup\{(\mu_{xy}(C),h,h/w)\}$}
}
\Return $\mathcal{V}$;\;
\end{algorithm}

\subsubsection{Span Generation: Heuristic Baseline}
Candidate pole-to-pole spans are generated using a conservative heuristic to ensure high recall. Each pole is connected to neighboring poles within a maximum span length, forming an over-complete candidate graph. For each candidate pair, wire-class support is evaluated within an oriented corridor aligned with the pole-to-pole direction \cite{csit_powerline_detection,isprs2019_utility_poles}. A span is retained only if (i) wire evidence is sufficiently dense and continuous, enforced by a minimum of 50 wire points and non-zero support in each 10\% segment along the span, and (ii) the corridor does not intersect any other detected pole, removing connections that geometrically pass through intermediate poles. The surviving edges define the heuristic span graph.

\subsubsection{Span Generation: Model-based}
We generate candidate pole pairs using a local neighborhood search constrained by a maximum span length (default \SI{200}{\meter}). Each candidate pair is then evaluated by a lightweight ResNet-based binary classifier \citep{he2016resnet} that operates on a rasterized 2D top view representation derived from the LiDAR point cloud, predicting whether a physical conductor span connects the two poles.

\paragraph{Candidate Pair Generation}
Poles are connected using a $k$-NN graph ($k{=}5$), with edges discarded if the pole-to-pole distance exceeds the maximum span length. This step intentionally over-generates possible connections to ensure high recall before span validation. (\autoref{fig:span_classified})

\paragraph{LiDAR-to-Image Representation}
For each candidate edge, an oriented rectangular corridor is extracted around the pole-to-pole segment. The corridor buffer is set adaptively as
\begin{equation}
b = \min(50\,\mathrm{m},\ 0.2\,d),
\end{equation}
where $d$ is the pole-to-pole distance (in meters). Wire and pole-class points within the corridor are rasterized at \SI{0.5}{\meter} resolution using a maximum-$Z$ projection \cite{lidar_dsm_projection} to preserve conductor geometry.

\begin{figure}[ht]
    \centering
    \includegraphics[width=0.8\linewidth]{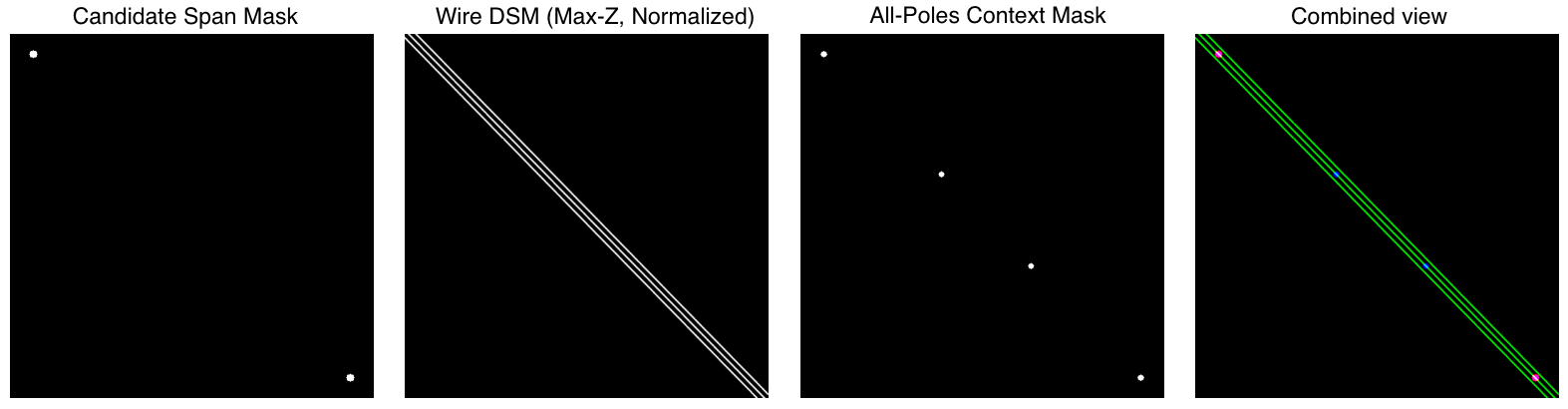}
    \caption{Multi-channel raster representation used for candidate span classification}
    \label{fig:model_input}
\end{figure}

\paragraph{Model Input}
Each candidate is represented as a $512 \times 512 \times 3$ raster image (\autoref{fig:model_input}) : (1) a candidate pole-pair mask encoding the endpoints and corridor geometry, (2) a min--max normalized wire DSM (max-$Z$ projection), and (3) an all-poles mask providing local network context. Binary masks are morphologically dilated to improve visibility under sparse returns and mitigate rasterization gaps. This 2D top view encoding captures the largely planar XY connectivity with height cues, enabling efficient CNN inference \cite{lawin2017deepprojection} while avoiding expensive 3D processing. 

\begin{figure}[ht]
    \centering
    \includegraphics[width= 0.8 \linewidth]{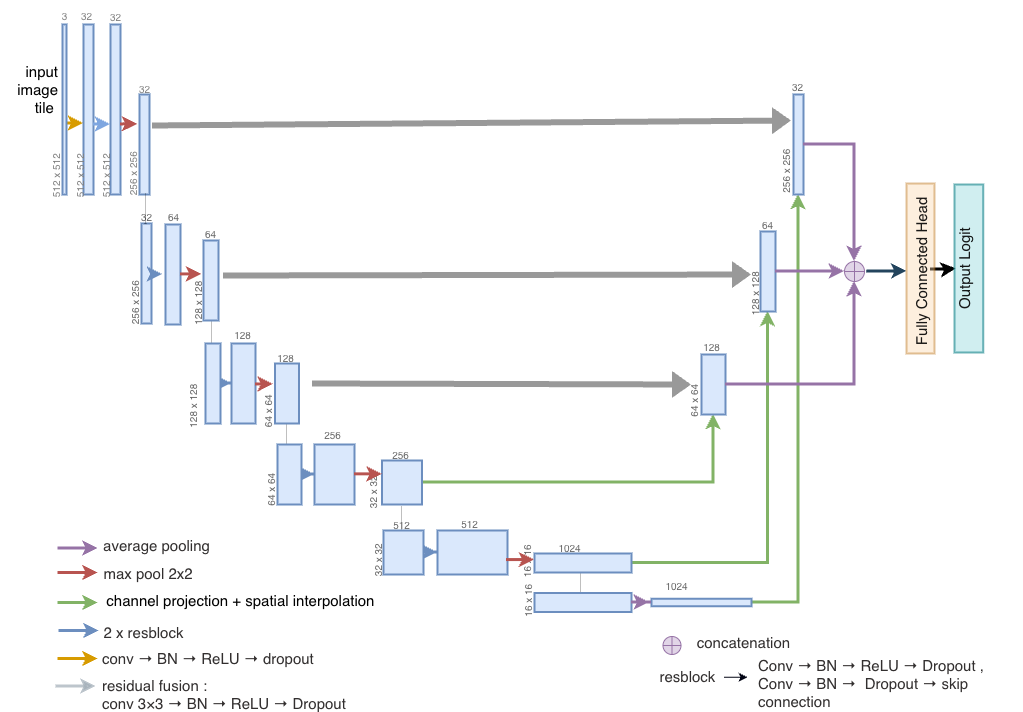}
    \caption{ResNet-based fusion classifier for span connectivity prediction}
    \label{fig:span_model}
\end{figure}

\begin{figure}[htbp]
    \centering
    \includegraphics[width=0.7\linewidth]{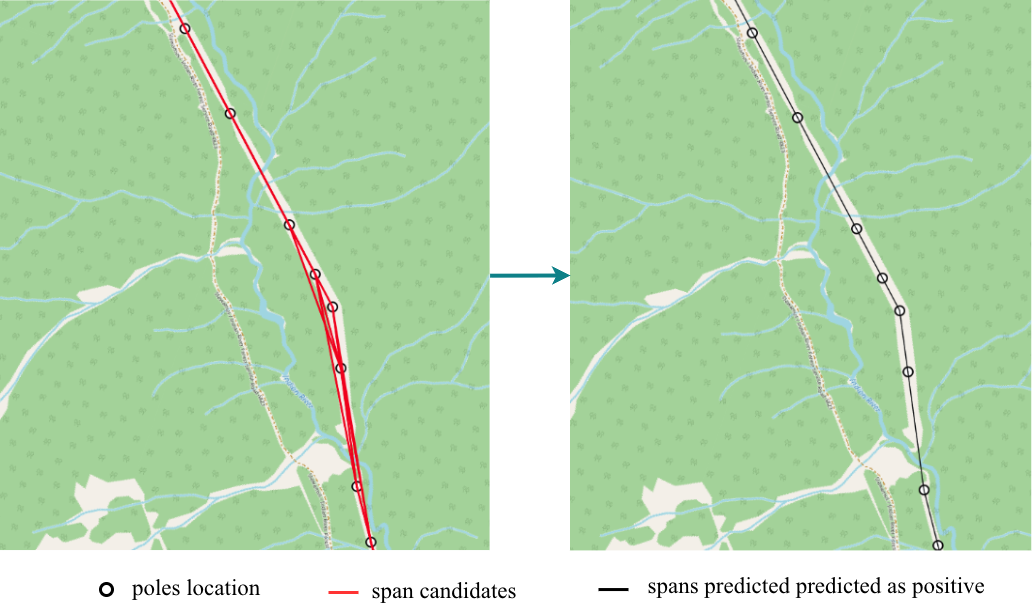}
    \caption{Candidate pole–pole spans (left) and predicted connected spans (right), yielding a consistent network topology
    }
    \label{fig:span_classified}
\end{figure}

\paragraph{Span Classification Network}
The classifier uses a custom ResNet model with multiscale feature fusion \cite{lin2017featurepyramid} to combine fine pole/wire cues with broader corridor context, improving robustness to scale variation and weak wire evidence. Candidate edges predicted as positive are aggregated to form the inferred span graph (\autoref{fig:span_model}). \autoref{tab:training_config} lists the training configurations for recreation of results

\paragraph{Wire Evidence Validation}
To improve robustness under segmentation noise and intermittent wire returns \cite{powerline_lidar_review}, predicted spans are post-validated using wire-class points inside the corridor by enforcing: (i) a minimum wire point count, (ii) continuity along the span to reject long gaps, and (iii) endpoint consistency near both poles. Spans failing these checks are flagged for downstream handling.

\begin{table}[ht]
\centering
\caption{Wire Classifier Training Configuration}
\label{tab:training_config}
\renewcommand{\arraystretch}{1.1}
\setlength{\tabcolsep}{5pt}
\begin{tabular}{p{0.30\linewidth} p{0.6\linewidth}}
\toprule
\textbf{Setting} & \textbf{Value} \\
\midrule
Input & $512 \times 512 \times 3$ (pole-pair mask, wire DSM, pole DSM) \\
Training data & 40k samples (10k positive and 30k negative) \\
Model/Params & Custom CNN with ResNet and feature fusion, 45.1M \\
Optimizer & AdamW ($\mathrm{lr}=10^{-4}$, $\mathrm{wd}=10^{-6}$) \\
Loss & Focal Loss ($\gamma=2.0$), $(\alpha_{\text{pos}},\alpha_{\text{neg}})=(0.73,0.27)$ \\

Augmentation & Rotation (positives only, $0^\circ$–$180^\circ$) \\

Split & 65/15/10/10 (train/val/test/holdout) \\
\bottomrule
\end{tabular}
\end{table}

\subsection{Stage 3: Span-Level Wire Metadata Extraction}
After topology reconstruction, we extract physical metadata for each validated pole-to-pole span. This is the outer boundary for the collection of wires going from one pole to another (\autoref{fig:outerbound}).

\subsubsection{Span Cropping from LiDAR}
For a span with endpoints $\mathbf{p}_1,\mathbf{p}_2$, we form a narrow oriented bounding box around the span segment (e.g., \SI{2}{\meter} width). Intersecting LAS tiles are queried, and wire- and ground-class points within the box are extracted and merged to obtain a span-level cropped point set.

\begin{figure}[ht]
    \centering
    \includegraphics[width=0.57 \linewidth]{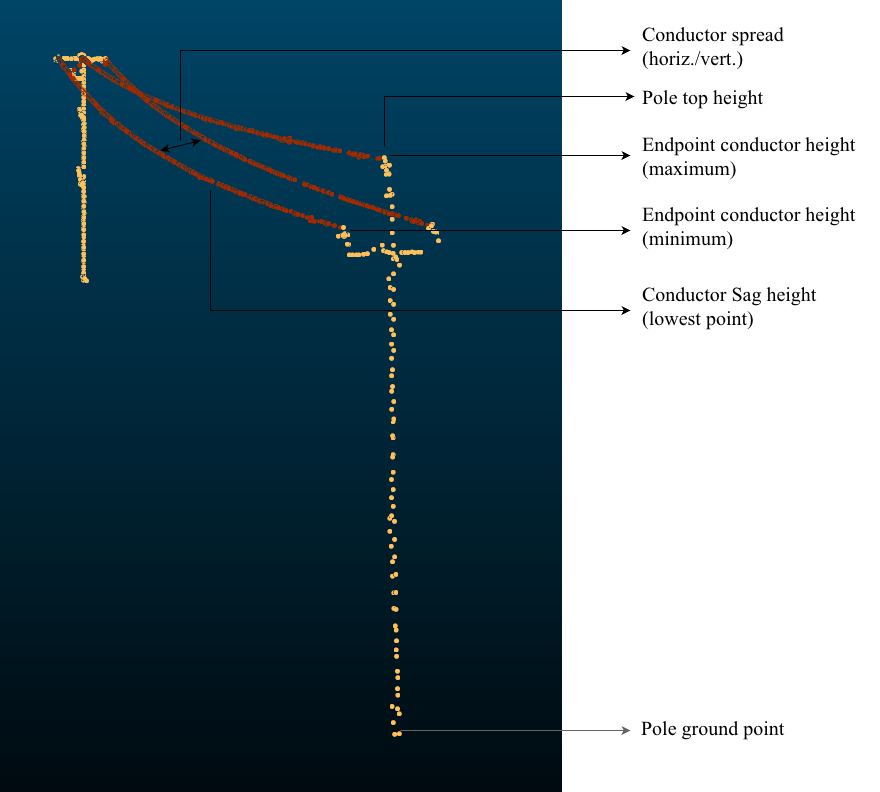}
    \caption{Attributes of span-level metadata extraction}
    \label{fig:span_labels_example}
\end{figure}

\subsubsection{Metadata Computation (\autoref{fig:span_labels_example})}
From the cropped span points (wire, pole and ground class only), we compute:
\begin{itemize}
    \item \textbf{Endpoint conductor heights:} robust estimates of conductor height near each endpoint (pole location) using percentile-based statistics.
    \item \textbf{Ground elevations:} ground height at both endpoints and near the span’s lowest-point neighborhood.
    \item \textbf{Sag-related features:} height and location of the lowest conductor point along the span.
    \item \textbf{Conductor arrangement and width:} number of conductors, arrangement inferred from horizontal/vertical spread, and endpoint wire width.
\end{itemize}

\section{Evaluation Data}
\autoref{tab:datasets} summarizes the datasets used for evaluation, many of which are proprietary. Metrics are reported consistently across all datasets and acquisition variations. For each dataset, spans were manually annotated by extracting wires and generating a reference shapefile containing span endpoints, conductor arrangement, sag-related attributes, and horizontal spread.

\begin{table}[htbp]
\centering
\caption{Datasets used for evaluation and benchmarking}
\label{tab:datasets}
\footnotesize
\setlength{\tabcolsep}{3pt}
\renewcommand{\arraystretch}{1.1}
\begin{tabular}{@{}l c p{1.8cm} r r r r@{}}
\toprule
\textbf{ID} & \textbf{Reg.} & \textbf{Scene} &
\textbf{Overall} & \textbf{Pole} & \textbf{Wire} & \textbf{Spans} \\
 &  &  & \textbf{PPM} & \textbf{PPM} & \textbf{PPM} & \textbf{Eval} \\
\midrule
A (2024) & US & For./Urb./Rur. & 16  & 12 & 8  & 3118 \\
B (2024) & ES & Urb./Sub.     & 40  & 20 & 10 & 401  \\
C (2023) & CA & Rur./For.     & 120 & 60 & 10 & 985  \\
DALES \citep{varney2020dales} & CA & Urb./Sub./Rur. & 50 & 20 & 12 & 4000 \\
\bottomrule
\end{tabular}

\vspace{0.5ex}
\footnotesize{\textit{For.=Forested, Urb.=Urban, Sub.=Suburban, Rur.=Rural.}}
\end{table}

		


\section{Results}

Evaluation metrics are tracked across different segments of the methodology.

\subsection{Point cloud segmentation}
\autoref{tab:ground_wires_poles_acc} denotes the performance on a per-class basis.

\begin{table}[htbp]
\centering
\caption{Per-class segmentation IoU for ground, wires, and poles across datasets}
\label{tab:ground_wires_poles_acc}
\renewcommand{\arraystretch}{1.1}
\begin{tabular}{lccc}
\toprule
\textbf{Dataset} & \textbf{Ground IoU} & \textbf{Wires IoU} & \textbf{Poles IoU} \\
\midrule
DALES                  & 0.98 & 0.77 & 0.65 \\
Other datasets & 0.97 & 0.91 & 0.63 \\
\bottomrule
\end{tabular}
\end{table}


\subsection{Topology Reconstruction Performance}
\begin{figure}[ht]
    \centering
    \includegraphics[width=0.7 \linewidth]{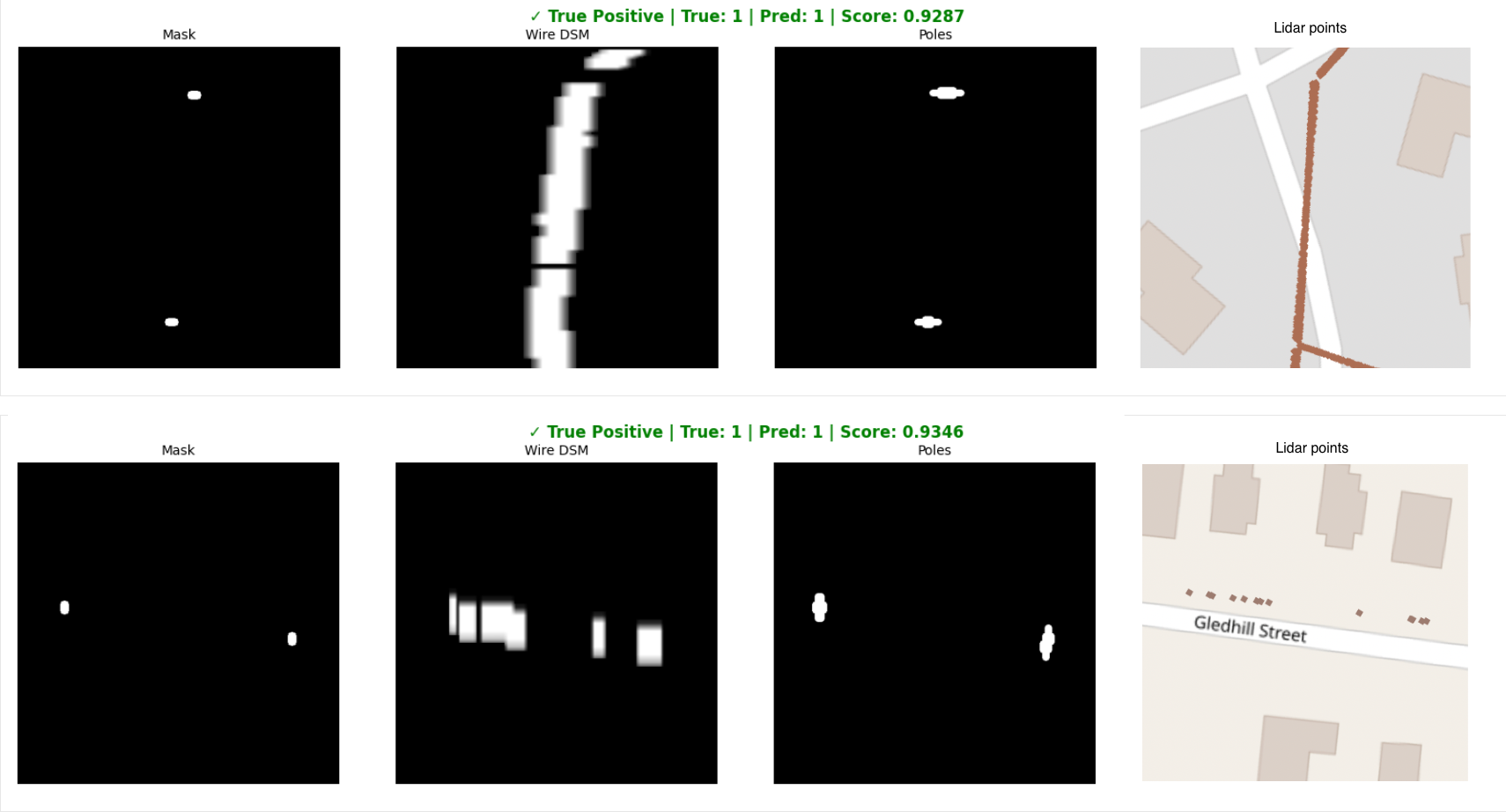}
    \caption{Sample input and outputs for the wire detection model}
    \label{fig:missing_wire_examples}
\end{figure}

Manually annotated spans were used as ground truth to evaluate (i) heuristic modeling and (ii) the proposed model-based span detection approach. \autoref{tab:topo_pr} reports precision and recall for both methods. A predicted span is correct if it connects the same pole pair as ground truth, where poles are matched within \SI{1}{\meter}. \autoref{fig:missing_wire_examples} illustrates cases where the proposed model correctly recovers spans despite sparse or largely occluded wire returns.

\begin{table}[ht]
	\centering
	\caption{Topology reconstruction performance (span-level).}
	\label{tab:topo_pr}
	\begin{tabular}{@{}lcc@{}}
		\toprule
		\textbf{Method} & \textbf{Precision} & \textbf{Recall} \\
		\midrule
		Heuristic baseline & 0.88 & 0.89 \\
		Model-based & 0.95 & 0.97 \\
		\bottomrule
	\end{tabular}
\end{table}

\subsection{Metadata Accuracy}
\autoref{tab:errors} reports mean absolute error (MAE) and root mean squared error (RMSE) for span-level attributes. The lowest-point estimate can exhibit higher error because aerial LiDAR often undersamples the lower conductors; occlusions and acquisition geometry lead to missing or sparse returns near the sag region, making the minimum-height measurement less reliable.

\begin{table}[ht]
\centering
\caption{Errors in span-level physical attribute estimation compared to manually labeled references (cm)}
\label{tab:errors}
\footnotesize
\setlength{\tabcolsep}{5pt}
\renewcommand{\arraystretch}{1.15}
\begin{tabular}{@{}p{4.8cm} r r@{}}
\toprule
\textbf{Attribute} & \textbf{MAE} & \textbf{RMSE} \\
\midrule
Endpoint conductor height (min) & 30.0 & 49.0 \\
Endpoint conductor height (max) & 19.25 & 45.5 \\
Conductor Sag height (lowest point) & 18.0 & 42.0 \\
Conductor spread (horiz./vert.) & 35.0 & 64.0 \\
\bottomrule
\end{tabular}
\end{table}

\section{Conclusion}


This paper presents an end-to-end framework for reconstructing overhead power network topology and extracting span-level physical metadata from large-scale aerial LiDAR. The pipeline covers the full workflow, from semantic segmentation of point clouds, through robust pole instance extraction and candidate span generation, to learning-assisted span validation and detailed wire metadata computation, resulting in a physically consistent pole-to-pole network representation suitable for engineering and GIS applications.

By combining conservative heuristic span generation with a lightweight ResNet-based classifier on compact 2D pole–wire evidence maps, the approach improves connectivity recovery under dense layouts and sparse or occluded wire observations while remaining efficient at scale. Experiments across diverse aerial LiDAR datasets demonstrate higher topology completeness and decimeter-level accuracy for key span attributes, enabling reliable end-to-end power network reconstruction.


\small
\bibliographystyle{IEEEtranN}
\bibliography{references}
\end{document}